\documentclass[journal]{IEEEtran}

\usepackage{framed,multirow}
\usepackage{amssymb}
\usepackage[numbers]{natbib}
\setcitestyle{numbers}
\usepackage{amsfonts}
\usepackage{amsmath}
\usepackage{latexsym}

\usepackage{url}
\usepackage{xcolor}
\definecolor{newcolor}{rgb}{.8,.349,.1}

\usepackage{hyperref}

\usepackage[switch,pagewise]{lineno} 
\usepackage{balance}
\usepackage{verbatim}
\usepackage{textcomp}
\usepackage{stfloats}
\usepackage{svg}
\usepackage{pifont}
\newcommand{\ch}{\ding{51}}%
\newcommand{\crs}{\ding{55}}%
\usepackage{rotating}
\title{\bfseries{A Phenomenological Approach to \\ Interactive Knot Diagrams}}
\author{Lennart Finke \footnote{} \thanks{Lennart Finke, with the Georg-August-Universität Göttingen, is the corresponding author.  E-mail: \href{mailto:l.finke@stud.uni-goettingen.de}{l.finke@stud.uni-goettingen.de}} \and Edmund  Weitz \thanks{Edmund Weitz is with the Hochschule für Angewandte Wissenschaften Hamburg. E-mail: \href{mailto:edmund.weitz@haw-hamburg.de }{edmund.weitz@haw-hamburg.de}.} \thanks{This work is submitted to the IEEE for possible publication. Copyright may be transferred without notice, after which this version may no longer be accessible.}}
\date{January 2024}

\begin{document}
\maketitle

\begin{abstract}
Knot diagrams are among the most common visual tools in topology. Computer programs now make it possible to draw, manipulate and render them digitally, which proves to be useful in knot theory teaching and research. Still, an openly available tool to manipulate knot diagrams in a real-time, interactive way is yet to be developed. We introduce a method of operating on the  geometry of the knot diagram itself without any underlying three-dimensional structure that can underpin such an application. This allows us to directly interact with vector graphics knot diagrams while at the same time computing knot invariants in ways proposed by previous work. An implementation of this method is provided.

\end{abstract}


\section{Introduction} \label{introduction}
\IEEEPARstart{A}{lthough} knots themselves are 1-manifolds with three-dimensional ambient space $\mathbb R^3$ or $S^3$, they have always been made sense of through their two-dimensional projections since the very first treatments by \textsc{Vandermonde} (1771) \cite{vandermonde1771remarques} or in unpublished work by \textsc{Gauss}. Classification of knots has been done by tabulating representative knot diagrams, most notably by \textsc{Rolfsen} (Appendix in \cite{rolfsen2003knots}, 2003), and the practice remains relevant as not only knot theorists but recently also applied scientists (for instance biologists \cite{brasher2013new}) take an interest in knot tables. Manipulating knot diagrams and recognising equivalent knots is often crucial, yet difficult to do mentally or on paper, as for example demonstrated by the case of the Perko Pair \cite{perko1974classification}. These difficulties make it desirable to employ digital tools for rendering, drawing and manipulating knot diagrams, facilitating a deeper understanding of knots.
Furthermore, teaching and learning knot theory can be aided greatly with interactive tools that make use of the visual, even haptic component of knots via diagrams. Finally, typesetting pretty knots with \LaTeX\, or 3D rendering software is often cumbersome and time consuming, so an intuitive way to draw as well as manipulate the geometry of knot diagrams in the same application with direct feedback has the potential to save a lot of time for researchers and educators. 

\begin{figure}
    \centering
    \includesvg[width=0.6\columnwidth]{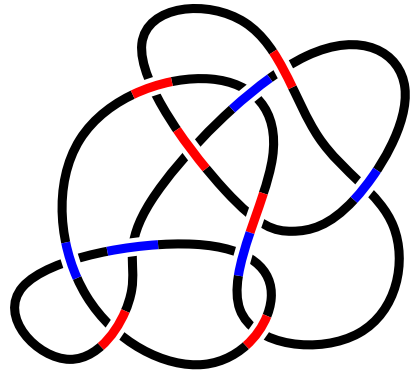}
    \caption{A knot diagram produced by our implementation. Intersections are coloured according to their over-under-boolean.}
\end{figure}

We aim to close the gap between these demands and existing solutions by introducing a method to handle both the geometry and topology of a knot diagram simultaneously through fluid motions facilitated by the user. A summary of previous applications is followed by an account of our proposal, then a description and evaluation of our openly available implementation and finally a proposal for possible future advancements.

\section{Precedent} \label{precedent}
We give an account of previous work on treating knots and knot diagrams computationally insofar as it pertains to our goal of interactively simulating knot diagrams.

Much theoretical and practical work has been done on computing a knot's or link's topology. The most straightforward and universal way to render knots on a computer, although we will explore a different avenue, is to maintain a three-dimensional geometry. The foundational work in this domain is \texttt{KnotPlot} \cite{scharein2002interactive} by \textsc{Scharein}, which is able to produce visually pleasing pictures of knots by converting an underlying three-dimensional polygonal line to either a projected knot diagram or a cylindrically encompassing mesh. Because manipulating these structures on a screen is tricky, a range of specialised tools is available to generate or alter them, such as generation by a self-avoiding random walk or pseudo-physical spring model to smooth a knot. We are interested in a more direct way of interacting with knot diagrams as two-dimensional structures, so we now outline the corresponding literature. The most widely used one today is presumably \texttt{SnapPy} \cite{SnapPy} by \textsc{Culler} et al. (2017). \texttt{SnapPy} is a popular Python library that can compute a number of invariants such as the Alexander polynomial, the Goeritz polynomial, the Jones polynomial, the Seifert matrix and the Floer homology. It features a graphical mouse input for link diagrams comprised of straight lines.
Other tools providing more advanced invariants include KnotKit \cite{seed2016knotkit} by \textsc{Seed} (2016) and \texttt{kht++} \cite{khtpp} by \textsc{Zibrowius} (2023).
A notable instance of graphical input of link diagrams is the ``Knot-Like Objects'' software \cite{swenton2021klo} by \textsc{Swenton} (2021). Here, the link is drawn with the mouse pixel-by-pixel. The bitmap is subsequently interpreted and converted into a geometric and finally a topological object. This approach is also taken by the Knot Identification Tool (2013) \cite{kit1013horowitz} by \textsc{Horowitz} (2013), a first adaptation of knot software to the web.
This design is combined with several invariants and the ability to parse photographs by the web application KnotFolio \cite{miller2022knotfolio} by \textsc{Miller} (2022), which was a particular inspiration for our efforts.
Drawing a knot diagram for subsequent analysis is also the primary objective of \texttt{KnotSketch} \cite{costagliola2016knotsketch} by \textsc{Costagliola} et al. Of particular note for us is the functionality to smooth the resulting diagram for improved visuals with a topology-respecting algorithm designed for arbitrary Euler Diagrams \cite{7192693}, adapting \texttt{KnotPlot}'s conception of smoothing an existing knot.

KnotPad \cite{knotpad} by \textsc{Zhang} et al. (2012) provides a tool for intuitively manipulating knots with an innovative approach; the user selects a region and proposes a modification by dragging the region with the mouse. If the proposal is a single Reidemeister move or a planar isotopy (that is, no crossings were interchanged, deleted or created), then the move takes place. This can be verified through the crossings on selected components before and after. Though promising, the software is not openly available. A strategy that reverses this strategy is implemented by \textsc{Liu} et al. (2020) \cite{liu2020suggestive} with the specific goal of finding knot diagrams with the fewest possible crossings. Their system recommends discrete steps of valid Reidemeister moves based on the diagram's Gauss code, which can then be carried out by the user through a raster drawing of the modification. Furthermore, the user can select between multiple rendering styles such as linked-node or a pseudo-three-dimensional rendering. Particular care for visual appeal is also given by \textsc{Eades} et al. (2023) in \texttt{CelticGraph} \cite{eades2023celticgraph}, a tool for creating artistic renderings of Celtic knots. Relevant for us is their use of cubic Bézier curves, since we will find them appropriate for our interactive vector graphics knot diagrams. A sensible alternative to using polynomial curves is however proposed by \textsc{Kindermann} et al. (2017) \cite{kindermann2017lombardi}. They study Lombardi drawings and, in particular, prove that every knot diagram is equivalent to a plane 2-Lombardi drawing, meaning that intersections are vertices connected by edges composed of two circular arcs, which are also visually pleasing and provide a guaranteed visual clarity. While a construction out of circles is somewhat at odds with our goal of interactivity, it could be an option for a final visual touch-up in the future.

\section{Method}
We describe a way to treat a knot diagram as a geometrical interactive object while keeping track of its topology throughout user manipulation. While representing the knot as a three-dimensional structure and rendering a knot diagram based on this would be viable, we want to acknowledge the success of vector graphics programs and let the user interact with the diagram itself as a self-intersecting curve. For this purpose, we render it to a viewport from which the user can see and interact with the knot diagram simultaneously, through both the mouse, keyboards and buttons that provide additional functionality. After a time window of receiving input, the calculation that preserves the underlying topology and creates the graphics for the next time window takes place. We call one such iteration a frame.
The knot diagram is represented as a closed, continuous path 
\begin{equation}
    \gamma(t): [0,1] \to \mathbb R^2, \gamma(0)=\gamma(1)
\end{equation}
consisting of several Bézier curves with shared start and end points. Computationally cheap algorithms exist to check for their intersections, which we use to calculate the (self-) intersections of the path at every frame. We store a boolean array that represents whether the first or the second location on the curve, as supplied by the intersection detection, goes above or below. More precisely, given an intersection $t_i<s_i$ with $\gamma(t_i)=\gamma(s_i)$, the $i$th boolean decides whether the segment around $t_i$ or $s_i$ goes above. This should take not much more than $30\,\text{ms}$ for perceived fluent motion.

\subsection{Interaction}
We allow the user to manipulate the geometry of the curve as follows. 
Clicking on the connection point between two neighbouring Bézier curves allows dragging it around. Instead of simply moving the control point to the location of the mouse, we only allow it to move below a certain small distance every frame to ensure that even a quick jerk of the mouse on a slow machine does not confuse the algorithm determining which intersections have moved where. A click \texttt{+shift} on a connection point deletes it and merges the neighbouring Bézier curves, whereas \texttt{+control} enforces $C^1$ continuity at the point by aligning the incoming tangents at the point to be equal and opposite, providing a smoothing effect. A click on the curve itself splits it in two by creating two curves where one was before. Control points of the Bézier curves can also be moved with click and drag, changing the shape of the path just the same. Specific to knots is that clicking on the region around an intersection (whose specific length will concern us below) flips the boolean value determining which strand goes above.

\begin{figure}
    \centering
    \includesvg[width=0.7\columnwidth]{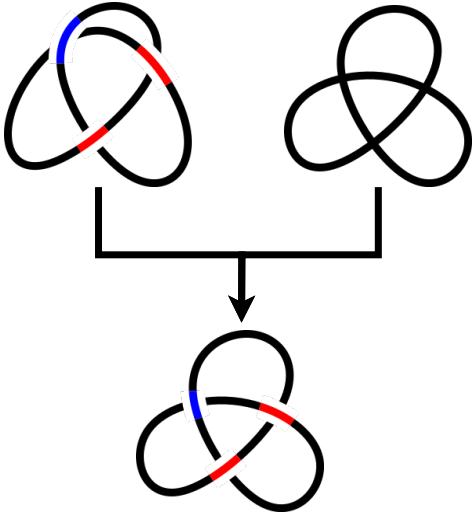}
    \caption{Inferring crossing booleans from a previous trefoil (top left) and a new curve (top right) manipulated by the user to a new trefoil (bottom). Knot diagrams were produced by our program.}
\end{figure}

Now the crucial step --- our program is supplied with a new Bézier path each frame after an interaction and is tasked with finding a boolean crossing array such that the path and crossings give a knot diagram that corresponds to a knot that is homotopic to the knot from the last frame. Sometimes, this is impossible, in which case the program should detect and perhaps report that to the user, for example by reverting the move.

Firstly, we must note that it is a well-known open question of knot theory whether it is possible to recognize if two knots are homotopic in polynomial time, or even if one is homotopic to the unknot. Nonetheless it might be tractable to take this approach as impressive progress into knot equivalence algorithms continues to be made, see for example the quasi-polynomial time bound algorithm given by \textsc{Lackenby} (2015) \cite{lackenby}. As of now though, no practical algorithm is in sight and we have to be inventive in using additional information to speed up computations enough to be feasible in a real-time interactive setting. 

One conceptually simple heuristic to find whether a boolean array is suitable is computing knot invariants of the new and old knot. If one of them disagrees between the two, they cannot be isotopic. However, this would only provide guarantees for correctly manipulating knots up to a certain crossing number, but certainly not in general. Additionally, it is computationally expensive to do this every frame. It might instead be conducive to do this only after some discontinuous action.

However, the most important manipulations of the path a user might make are continuous, which we can use to our advantage. In particular, we can then bijectively map the crossings from the previous frame $\mathcal K = \{(t_1,s_1), \dots, (t_n,s_n)\}$ to the crossings of the next frame $\mathcal K' = \{(t'_1,s'_1), \dots, (t'_m,s'_m)\}$ based on some notion of minimal distance. To precisely refer to these points $\gamma(t_i) = \gamma(s_i)$ we additionally adopt the convention $t_i < s_i$ and assume there are no triple or more intersections on the same point in space. Since intersections can also appear or vanish between frames, we further introduce the possibility of mapping to and from virtual points $\star := \{\star_j \mid 1\leq j \leq |\mathcal K'| - |\mathcal K|\}$ and $\dagger := \{\dagger_j \mid 1\leq j \leq |\mathcal K| - |\mathcal K'|\}$.
We are confronted with a simple optimal transport problem; firstly we choose a sensible metric $d_\gamma: [0,1]^2 \times [0,1]^2 \to\mathbb R$ that is possibly based on the previous path $\gamma$. We then assign a finite, but arbitrarily high cost for creating or deleting points by setting $d_\gamma(\star_j, \cdot) := d_\gamma(\cdot, \dagger_i) := C$, where $\cdot$ is any value and $C$ is larger than the maximum of $d_\gamma$ over $\mathcal K \times \mathcal K'$. Now we find a bijective map $\varphi^\star: \mathcal K \cup \star \to \mathcal K' \cup \dagger$ that moves the intersections as little as possible with respect to it

\begin{equation}
    \varphi^* := \underset{\varphi \in \text{Bij}(\mathcal K \cup \star , \mathcal K' \cup \dagger)}{\arg \min}\left\{ \sum_{x\in\mathcal K \cup \star} d_\gamma(x, \varphi(x))\right\},
\end{equation}
which is used to copy the over-under boolean from an intersection $(t_i, s_i)$ to $\varphi^\star((t_i, s_i))$. Instead of requiring a metric, it can also be valid to use a pre-metric for $d_\gamma$, that is loosen the constraint $d_\gamma((t_i,s_i), (t'_i,s'_i)) = 0 \Longrightarrow (t_i,s_i) = (t'_i,s'_i)$, although as we will see below this can cause poor results. Mapping from $\star$ or to $\dagger$ alerts us that crossings have been deleted or created. We can only allow this by way of a Reidemeister move, so we check whether a valid move could be recognized by a procedure described below. In case the move was deemed to be valid, we copy over the boolean from $(t_i, s_i)$ to $\varphi((t_i, s_i))$ or assign it a new boolean dictated through a Reidemeister move if it was created anew.

Because this process needs to be carried out at least several times a second for a fluent interaction, we might want to solve this problem only approximately and simply set \begin{equation}
    \varphi^\star((t_i,s_i)) = \underset{(t_i',s_i') \in \mathcal K'}{\arg \min}\{d_\gamma((t_i,s_i), (t_i',s_i'))\},
\end{equation}
if $|\mathcal K| \leq |\mathcal K'|$ and defining via the inverse
\begin{align}
    &(\varphi^\star)^{-1}((t'_i,s'_i)) := \underset{(t_i,s_i) \in \mathcal K}{\arg \min}\{d_\gamma((t_i,s_i), (t_i',s_i'))\}
\end{align}
if $|\mathcal K| > |\mathcal K'|$. To formally ensure that $\varphi^\star$ is well defined, we additionally choose the lowest index $i$ in the $\arg \min$ when two distances are equal. All unmapped intersections are recorded and then understood to be reached from $\star, \dagger$ respectively.

What remains to be discussed is the choice of distance $d_\gamma$. Intuitively, it might seem fine to simply use the Euclidian distance between two intersections in space

\begin{equation}
    \begin{aligned}d_\gamma((t,s), (t',s')) &= ||\gamma(t') - \gamma(t)||_2 \\ &= ||\gamma(s') - \gamma(s)||_2\end{aligned}
\end{equation}
and indeed this works in most situations. The problem arises when the user attempts to make a legal Reidemeister 3 move and the three time point pairs coalesce into a single intersection in space. In the exact frame the move occurs, the three pairs are so close to each other in ambient space that they will end up matched to essentially a random other one. If one is bent on using spatial distances, the (almost-everywhere existent) tangent vector of the strand going over may be included in $d_\gamma$. Instead we can more aptly use the fact that spatially intersecting points remain far apart in terms of parameter space, while being close to their previous frame. Thus we arrive at a periodic curve time distance

\begin{equation}
    \begin{aligned}
    d_\gamma((t,s), (t',s')) = &\sin(\pi \cdot |t-t'|) + \sin(\pi \cdot |s-s'|).
\end{aligned}
\end{equation}

It ensures that both intersections viewed as angles on the curve will not move by much and thus also stay in the same location. Instead of $\sin(\pi \cdot x)$, any other continuous function strictly increasing on $[0,0.5)$ and symmetric about $0.5$ will do as well. When executing the described procedure with this metric, intersections can be correctly mapped not just through planar isotopy moves but also legal Reidemeister 1, 2 \& 3 moves. It does however not safeguard against attempted Reidemeister-like illegal moves, which we will treat now.

\begin{figure}
    \centering
    \includesvg[width=0.6\columnwidth]{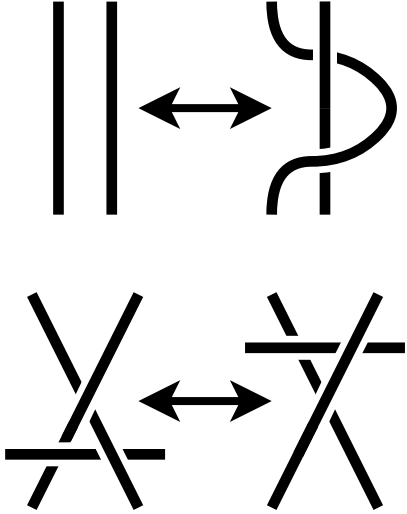}
    
    \caption{The \textit{Reidemeister Crimes}. These are the illegal moves that we need to revert once the user attempts them. Above is a wrong version of Reidemeister 2, below a wrong version of Reidemeister 3 where the left side shows a tangle. Any combination of over-under-boolean on the bottom right would be illegal.}
\end{figure}
\begin{figure}
    \centering
    \includesvg[width=0.9\columnwidth]{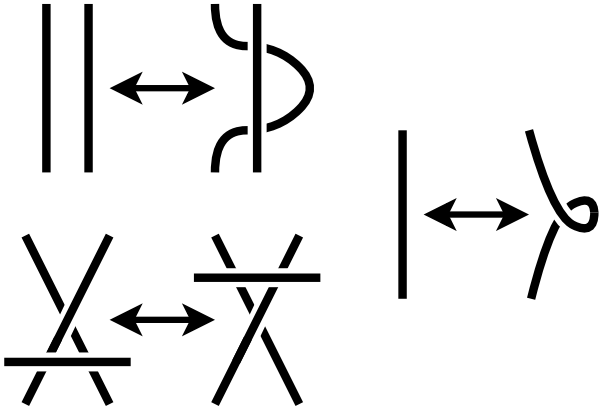}
    
    \caption{The Reidemeister Moves. These preserve the topology, so we allow an operation if it can be identified as one of them.}
\end{figure}

In order to prevent wrong versions of Reidemeister moves, we have to look at them one by one: 

\begin{enumerate}
    \item The first Reidemeister move cannot go wrong. If a single intersection is created, we arbitrarily assign it a boolean \texttt{false}. If it is deleted, we forget about its boolean.
    \item The second is slightly trickier. If two intersections are created, we assign both the same boolean. To anticipate the intuition of the user, we put $t_j$ over $s_j$ or vice versa depending on which is closer to the time point selected by the user. This creates an effect of picking a part of string and pulling it over another. A special case can occur when the point that is pulled over is exactly the location $\gamma(0)=\gamma(1)$, so we have to use the periodic distance $d_\gamma$ once again to determine which time point is closer to the selected one. If in turn two intersections are deleted, we have to check whether the two adjacent (in terms of $d_\gamma$) pairs $(t_i, s_i), (t_j, s_j)$ have the same boolean. The move is only executed when either both $t_i, t_j$ or both $s_i, s_j$ go over, that is when both have the same boolean. Otherwise, the move might not have been a homotopy. 
    \item The third Reidemeister move is also the most difficult. For legal moves, we already took care of choosing the right booleans for the frame after the move in the considerations above, so we need only detect when three intersections $(t_1,s_1), (t_2,s_2), (t_3, s_3)$ involved in the move form a tangle. A tangle is formed when none of the three strands goes over the other two. There are many ways to extract this from the data our application has access to, for example going through all possible combinations of two intersections. A spatial way to approach it would be to check if the dot product of the normalised tangent $\dot \gamma (t) \cdot \dot \gamma(s)$ of two of the intersection at the strands going over $t$ and $s$ approaches 1 as the intersections get closer, which has to be the case for one pair in a legal Reidemeister 3.
\end{enumerate}

Apart from dragging segments of the knot, it can also be useful to mirror, rotate, translate or reverse the curve. These operate on the whole of the knot diagram in a predictable way and consequently do not pose a challenge to our algorithm determining which intersections in the next frame belong to which in the last frame, but it is nonetheless important to consider them. If such a predictable discontinuous operation is performed on the Bézier curve path, one also needs to perform it on the data structure that keeps track of the intersection. When deleting and adding Bézier curves, we also must update the curve times of all intersections.

At the end of the process of mapping to the new intersections, we have to do an additional check to detect intersections $(t_i, s_i)$ that have looped once around the path by crossing $\gamma(0)=\gamma(1)$. They switch from $t_i < s_i$ to $s_i < t_i$, since the larger one becomes $s_i \approx 0$ in the frame after it loops around. We consequently have to switch the boolean value corresponding to this intersection. Detecting this is readily done by keeping track of which Bézier curve an interaction is located on and recording jumps in the Bézier curve index of an intersection that do not also coincide with a jump in space, $||\gamma(s_i) - \gamma(s_i')||_2 < \epsilon$, where the threshold $\epsilon$ depends on units of space. It would also be possible to let $\epsilon$ be a function of the time elapsed since the last frame, but since we have constrained the mouse movement anyway this will not be necessary.  

After the intersections have been matched successfully and the knot diagram for the next frame is thereby specified, we can proceed with calculating knot invariants and classification if needed. When doing this, it is especially useful if algorithms calculating an invariant only take in the intersection locations of the knot diagram and do not need further geometrical considerations. Taking the Alexander polynomial as an example, we can avoid having to use enclosed regions by following a slight variation of the original definition by \textsc{Alexander} (1928) \cite{alexander1928topological} phrased in terms of a diagram's arcs. A concise description is given by \textsc{Lopez} (2003) \cite{lopezalexander}.

In case an illegal move is detected, we revert the knot to its state before the user interaction. We can also alert the user as to what rule was violated, for example that an illegal Reidemeister 2 was performed. If the user action was of discrete nature, such as a deletion of a control point, and we detect a change in the diagram through a change in the Dowker-Thistlethwaite code, we can additionally warn them of this to remind them that isotopy could not be guaranteed.

\begin{figure}
    \centering
    \includegraphics[width=\columnwidth]{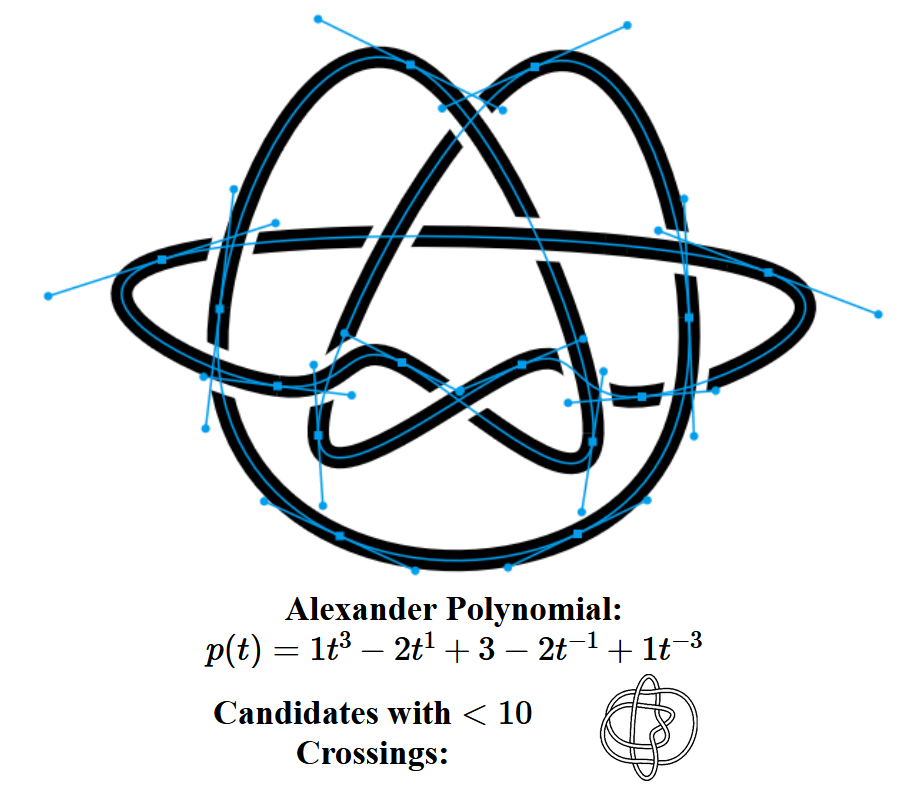}
    \caption{A screenshot of the implementation showing one representative of the Perko pair, pieced together by continuously differentiable Bézier curves with visible handles. Blue dots can be moved with the mouse. The Alexander polynomial correctly recognises the knot as entry $\texttt{1\_161}$ in the Rolfsen table\cite{rolfsen2003knots}. One application of the program could be to transform the diagram above into the representation given by the Knot Atlas \cite{knotatlas}.}
\end{figure}

\subsection{Visuals}

Finally, the resulting knot diagram needs to be presented back to the user through a graphical interface. This is not as straightforward as it might seem, because as opposed to other methods, our real time approach does not give any guarantees for the absence of cusps or for a limitation on the number of intersections in one area. Further, we have to deal with intersections between two Bézier curves, just as with self-intersections of a single Bézier curve. It can also be the case that an intersection is arbitrarily close to the connecting point between two Bézier curves.

We take inspiration from a \LaTeX-library for rendering knots, \texttt{spath3} \cite{spath3} by \textsc{Stacey} (2011). \texttt{spath3} segments the knots between crossings and draws strands that go above double; one iteration is thick and in the background colour, the other in regular thickness and has the colour of the knot. Thereby, the strand going under looks like it vanishes below the other before it reemerges on the other side. The trouble with using this for our purposes is that segmenting the knot in this way would create a separation between what the user interacts with and what they see. For instance, dragging the connection points would not be possible continuously, since every segment may then only contain at most one intersection. Taking a different approach and drawing twice on all of the segment despite multiple intersections would lead to doubly drawn strands overriding each other, so that depth information on which strand needs to be drawn first becomes necessary.

Our solution is to draw doubly only in a neighborhood around the intersection and along two neighboring Bézier curve segments along the path. We then need an additional consideration regarding the length of this small piece: Short pieces are good to not block other parts of the knot for other intersections or user interaction, long pieces are needed because they have to cover up the strand going under. A sensible heuristic to cover just as much of the bottom strand as needed is calculating the time-length radius of the doubly drawn piece $l$ by approximating both curves as straight lines and making due with finding the length along the upper strand from the intersection of the center of the two lines up to the intersection of the edges of the two lines including a gap width. Through two simple trigonometric calculations we obtain

\begin{align}
    \alpha &:= \frac{\pi}{2} \frac{|\dot \gamma(t) \cdot \dot \gamma(s)|}{||\dot \gamma(t)||_2 \cdot ||\dot \gamma(s)||_2}, \\
    l &:=  (w_s+w_g) \cdot \tan (\alpha) + w_s\cdot\sec(\alpha),
\end{align}
where $\sec$ is the inverse cosine, $\gamma$ is again the path, $w_s$ the stroke width, $w_g$ the desired gap width and $t,s$ the times of the path where it crosses itself. This formula assumes that the Bézier curve times are sampled adaptively, so that one unit in parameter space is one unit in real space. This is the case in our implementation, but if it is not, an appropriate rescaling should first take place. It can further be troublesome that $l$ is not bounded as a function of $\alpha$, in which case we recommend to fix an upper limit or limit the intersection to the neighbouring Bézier curves only. The doubly drawn piece is finally $\gamma([t - l, t + l])$ or $\gamma([s - l, s + l])$ depending on which strand is above, where the interval bounds are interpreted modulo 1. This ensures that perpendicular strands are only drawn doubly on a short segment, while more parallel crossings are also fully covered.

For pleasant visuals when finally exporting the knot, we provide a feature to keep the knot smooth by enforcing $C^1$ continuity. To enter this state, a centripetal Catmull-Rom spline \cite{catmull1974class} through the control points is calculated, which is however a discontinuous operation and cannot be guaranteed to preserve the topology. Another discontinuous operation we make available is straightening the knot diagram into a polygonal line. For both of these, we provide a warning if the Dowker-Thistlethwaite code has been changed from before the operation, which would be a necessary but not sufficient condition for a move that breaks the isotopy.

Simply smoothing the spline that makes up the knot is not a guarantee for an aesthetic diagram. The user is therefore left with the task of editing the Bézier curve -- a time-consuming process that prompts further automation. We are evidently faced with a problem that has many possible solutions as beauty is famously subjective. Nonetheless most will value a knot diagram exhibiting few bends, moderate curvature, equally sized enclosed areas, equally spaced crossings, intersections at right angles and symmetry. One could also ask for a sort of canonical knot diagram, such as one composed of the fewest possible Bézier curves (an open problem as far as we know), one with the most symmetries or a polygonal chain with the least bends.

One of the ways to approach this rather fuzzy problem consists in thinking topologically. We forget that we are trying to display a spline or polygonal line in the end and interpret the diagram as a bidirected graph, where the vertices are intersections and the edges are the arcs of the knot leading from one intersection to the next. Our goal is then to find a planar embedding that satisfies properties that lead to a pleasing diagram. We deem the approach of \texttt{SnapPy} \cite{SnapPy} to be the most appropriate in our case, amended by the note that the graph embedding literature is extensive and might harbor potential for improvement. In particular, Culler et al. notice that every vertex has degree four and apply an algorithm by \textsc{Tamassia} (1987) \cite{doi:10.1137/0216030} based on network flow methods that finds a geometric embedding with the minimal number of bends of those with edges only on the integer grid. The graph is then converted back to a polygonal chain with the crossing information stored in the vertices. The crossings are now equally spaced, they intersect at right angles and the enclosed areas are small. This provides if not a particularly striking beauty at least a tidy, canonical version of any given diagram. In our tool, where we used the corresponding module from \texttt{SnapPy} to showcase the utility of this approach, the polygonal chain can subsequently be smoothed and edited again.
Another conceivable sibling of this topological method would be to find a Lombardi drawing of the diagram, as described in Section \ref{precedent}, but computational methods are lacking as of now.

\section{Results \& Evaluation}

We provide an interactive online implementation called \texttt{Knottingham} at \url{https://fi-le.net/knottingham} with source code openly available at \url{https://github.com/file-acomplaint/knottingham} under an MIT license. It is written in JavaScript on a static \texttt{HTML} page and runs on all common web browsers. Although it is designed for the computer mouse, using the application from a mobile device is also possible, with limited features. For handling graphic display, drawing and detecting intersections, the JavaScript library \texttt{paper.js} \cite{lehni2011paper} by \textsc{Lehni} and \textsc{Puckey} (2011) is used. In particular, it handles the rendering of Bézier curves and uses adaptive sampling and the monomial basis together with a sort of Horner's method \cite{horner1819xxi} to alleviate the resulting numerical instability.

A comparison of the features of our implementation with other software is provided in table \ref{comparison}.

\smallskip

\begin{table}
    
    \centering
    \caption{A qualitative comparison between the features of related programs described in section \ref{precedent} and ours on the right.}
    
    \begin{tabular}{l|lllll} \label{comparison}
        & \begin{turn}{90}KnotFolio \cite{miller2022knotfolio}\end{turn}
        & \begin{turn}{90}KnotPad \cite{knotpad}\end{turn}
        & \begin{turn}{90}KLO \cite{swenton2021klo}\end{turn}
        & \begin{turn}{90}SnapPy \cite{SnapPy}\end{turn} 
        & \begin{turn}{90}Knottingham\end{turn} \\ \hline
        Freehand Drawing
        & \ch & \crs & \ch & \crs & \ch \\\hline
        Alexander Polynomial
        & \ch & \crs &  \ch & \ch & \ch \\\hline
        Other Invariants, \\ i.e. Floer Homology
        & \ch & \crs & \ch & \ch & \crs \\\hline
        {Diagram Manipulation}
        & \crs & \ch & \ch & \crs & \ch \\
        {\dots with Do/Undo}
        & \crs & \ch & \ch & \crs & \ch \\
        {\dots strictly by isotopy}
        & \crs & \ch & \ch & \crs & \crs \\
        {\dots fluently}
        & \crs & \ch & \crs & \crs & \ch \\
        {\dots in real time}
        & \crs & \crs & \crs & \crs & \ch \\\hline
        Exporting to \\ \texttt{SVG}, \texttt{JSON} \& \texttt{TikZ}
        & \crs & \crs & \crs & \crs & \ch \\
        {+ re-importing \texttt{JSON}}
        & - & - & - & - & \ch\\\hline
        {Rotating, Flipping}
        & - & - & - & - & \ch \\
        {Bézier Curve Controls}
        & - & - & - & - & \ch\\
        {Visual Tweaks}
        & - & - & - & - & \ch\\
        \hline
        \end{tabular}
    \label{tab:my_label}
\end{table}

\smallskip

With our implementation, it is possible to draw a knot, then manipulate it in real time and export to graphical output formats or a re-importable \texttt{JSON}. The Alexander polynomial is also computed, demonstrating that other more involved invariants are also possible. With enough invariants, one can then attempt a classification of the knot that was created by the user, by searching knot databases for matches in all invariants. We follow the approach of the Knot Identification Tool \cite{kit1013horowitz} and \texttt{KnotFolio}\cite{miller2022knotfolio}, doing precisely this kind of classification with the Alexander polynomial and the Rolfsen knot table \cite{rolfsen2003knots} on every frame. Gauss codes and Dowker-Thistlethwaite codes \cite{dowker1983classification} are also provided. We then display a representative diagram with a hyperlink to the corresponding entry in the Knot Atlas \cite{knotatlas}, an open encyclopedia run by \textsc{Bar-Natan} and \textsc{Morrison}. On top of this, we can detect illegal Reidemeister moves in fluent motion, at least given a sufficiently high frame rate. This combination of features makes our implementation not only novel, but also helpful in concrete use cases for researchers and students. 

We now address the shortcomings of our implementation. Most importantly, we cannot strictly guarantee that the knot is only transformed by homotopy. In practice, even on slow machines, the mapping works fine, but user discretion is necessary all the same; the system is only an aid for thinking through knot equivalences, not a replacement for it. For the time being we have made due with appropriate precautions and warnings, but we believe that one can amend this into a system that can step through knot equivalences with the certainty of a computer-assisted proof. Another inconvenience of the current version is that the over-under booleans cannot be exported to \texttt{TikZ} as it uses an involved system for numbering the intersections that would have to be reverse-engineered to be interoperable.

\section{Outlook}
We have introduced a method to represent knot diagrams through a path of Bézier curves and supplementary data that is used to preserve the topology of the underlying knot. We have further presented an implementation of this method and discussed its benefits as well as drawbacks. Based on this, several possible improvements suggest themselves for future work. Most importantly, the main idea of mapping an array of booleans to preserve the object's topology could be extended to link and braid diagrams. While analogous in principle, this might involve some additional considerations expanding on the above as well as software engineering challenges, suggesting itself as the subject of research to come.

Another area of significant potential is guaranteeing to allow provably only strict knot equivalences as described in the previous section, which is not realised here.

It might also be possible to significantly improve computational efficiency, leading to a better user experience even when handling larger knots on weak computers. As far as the implementation is concerned, a dedicated version for mobile use could further improve ease of use. More invariants and characteristics are also well within the realm of possibility, as the computation of the Alexander polynomial demonstrates. When implementing them, one would need to also consider how to compute quickly enough so as to maintain a real-time display or perhaps compute them only if prompted by the user. If more invariants are available, this would allow for classification of more knots, especially beyond the Rolfsen table. More extensive tables are already available for this, like for example KnotInfo \cite{knotinfo} maintained by \textsc{Livingston} et al. (2023).  

We have described a topological approach to computing a more aesthetically pleasing diagram from the one created by the user, but it must be noted that this runs somewhat contrary to our goal of a true interactive knot diagram. While we can guarantee that the diagram resulting from such a reconstruction through a topological graph represents the same knot as before, it may not be obvious to the user that this is indeed so and they may rightfully ask for a series of deformations towards this prettier shape. While we have not solved this ourselves, we nonetheless would like to suggest a blueprint in the form of a second approach, which we call physical. This time, we approximate our spline with a graph that is already embedded, so vertices with degree four at the intersections and periodically spaced vertices with degree two between them. We then take inspiration from \texttt{KnotPlot} and run a spring force model on this structure, automatically contracting excessively large areas in the diagram and smoothing out the arcs as the spring forces approach equilibrium. Finally, we may then reconstruct a polygonal chain from this embedded graph as we can keep track of the crossing information in the intersection vertices and optionally approximate the polygonal chain by a Bézier spline. Our experiments have not yielded satisfactory results with a naive spring model, so it appears as though the more recent literature about numerical methods on knots should be put to use. In particular, we suspect that the method of \textit{Repulsive Curves} \cite{Yu:2021:RC}, introduced for the purpose of untangling knots in three-dimensional space by \textsc{Yu} et al. (2021), can help us find a diagram with an evenly spaced layout of both intersections and the enclosed areas in a diagram through the notion of a non-locally determined force acting on the vertices. We are in the process of applying these and similar findings to our tool.

Since we have significantly improved upon the immediateness of feedback to the user, we can also recommend adopting a more pedagogical view and developing software for students or others who might not already have a scholarly interest in knots. This could involve visual upgrades like a pseudo-3D look, as seen in KnotPad \cite{knotpad}, or a more modern user interface.

We thus understand the potential for interactive knot software to be still high and  look forward to further developments.

\bibliographystyle{abbrvnat}
\bibliography{bibliography}

\end{document}